\documentclass[superscriptaddress,reprint,prb]{revtex4-1}
\pdfoutput=1
\usepackage{graphicx}
\usepackage[margin=0.7in]{geometry}
\usepackage{amsmath}
\usepackage{hyperref}
\usepackage{multirow}

\renewcommand{\Im}{\operatorname{Im}}

\newcommand{\sub}[1]{\ensuremath{_{\textsf{#1}}}}  
\newcommand{\RPIMSE}{Department of Materials Science and Engineering, Rensselaer Polytechnic Institute, Troy, NY, USA}
\newcommand{\RPIPhy}{Department of Physics, Applied Physics and Astronomy, Rensselaer Polytechnic Institute, Troy, NY, USA}
\newcommand{\HarvardFAS}{Faculty of Arts and Sciences, Harvard University, Cambridge, MA, USA}

\usepackage[usenames]{color}

\begin{document}

\title{Effects of Interlayer Coupling on Hot Carrier Dynamics\\in Graphene-derived van der Waals Heterostructures}

\author{Prineha Narang}\email{pnarang@fas.harvard.edu}\affiliation{\HarvardFAS}
\author{Litao Zhao}\affiliation{\RPIMSE}
\author{Steven Claybrook}\affiliation{\RPIPhy}
\author{Ravishankar Sundararaman}\email{sundar@rpi.edu}\affiliation{\RPIMSE}\affiliation{\RPIPhy}

\date{\today}

\begin{abstract}
Graphene exhibits promise as a plasmonic material with high mode confinement that could enable efficient hot carrier extraction.
We investigate the lifetimes and mean free paths of energetic carriers in free-standing graphene, graphite
and a heterostructure consisting of alternating graphene and hexagonal boron nitride layers using
\emph{ab initio} calculations of electron-electron and electron-phonon scattering in these materials.
We find that the extremely high lifetimes (3 ps) of low-energy carriers near the
Dirac point in graphene, which are a hundred times larger than that in noble metals,
are reduced by an order of magnitude due to inter-layer coupling in graphite,
but enhanced in the heterostructure due to phonon mode clamping.
However, these lifetimes drop precipitously with increasing carrier energy,
and are smaller than those in noble metals at energies exceeding 0.5 eV.
By analysing the contribution of different scattering mechanisms and inter-layer interactions,
we identify desirable spacer layer characteristics -- high dielectric constant and heavy atoms --
that could pave the way for plasmonic heterostructures with improved hot carrier transport.
\end{abstract}

\maketitle

Two-dimensional (2D) materials exhibit a diverse array of optical and electronic properties,
ranging from insulating hexagonal boron nitride and semiconducting transition metal dichalcogenides
to semimetallic graphene.\cite{RevModPhys.81.109, Novoselov666, Geim:2007ty, R.:2010qv, PhysRevB.76.073103}
Stacked 2D materials, or van der Waals (vdW) heterostructures,\cite{Xia:2014lq, Geim:2013ud, Jariwala:2016zp}
have generated considerable recent interest as designer plasmonic, photonic and optoelectronic materials.
Combining 2D layers in different arrangements makes it possible to realize a variety of new optical phenomena
and nanophotonic devices, covering spectral ranges from the microwave to the ultraviolet.\cite{Ci:2010rz,PhysRevB.80.245435, Bonaccorso:2010uo}

Simultaneously, the field of utilizing the energetic `hot' carriers generated by surface plasmon decay
for photodetection and solar energy conversion has grown rapidly.\cite{Brongersma:2015fk, Prineha:2016fk,Xia:2009dn}
Hot carrier extraction has also been demonstrated in graphene,\cite{Oum:2014uq, PhysRevLett.107.237401}
with experimental techniques such as pump-probe spectroscopy\cite{Norris}
and four-dimensional electron microscopy\cite{Zewail187, Zewail-UED-Graphene}
used to explore the energy relaxation dynamics.\cite{PhysRevLett.102.086809, Carbone2011}
However, these techniques conventionally provide indirect signatures
of the response of a large number of thermalizing carriers,\cite{Knorr2007}
and extensive theoretical modeling is necessary to extract information about the sub-picosecond
non-equilibrium carrier dynamics of interest.\cite{ Tielrooij:2013ek}

\begin{figure}
\includegraphics[width=\columnwidth]{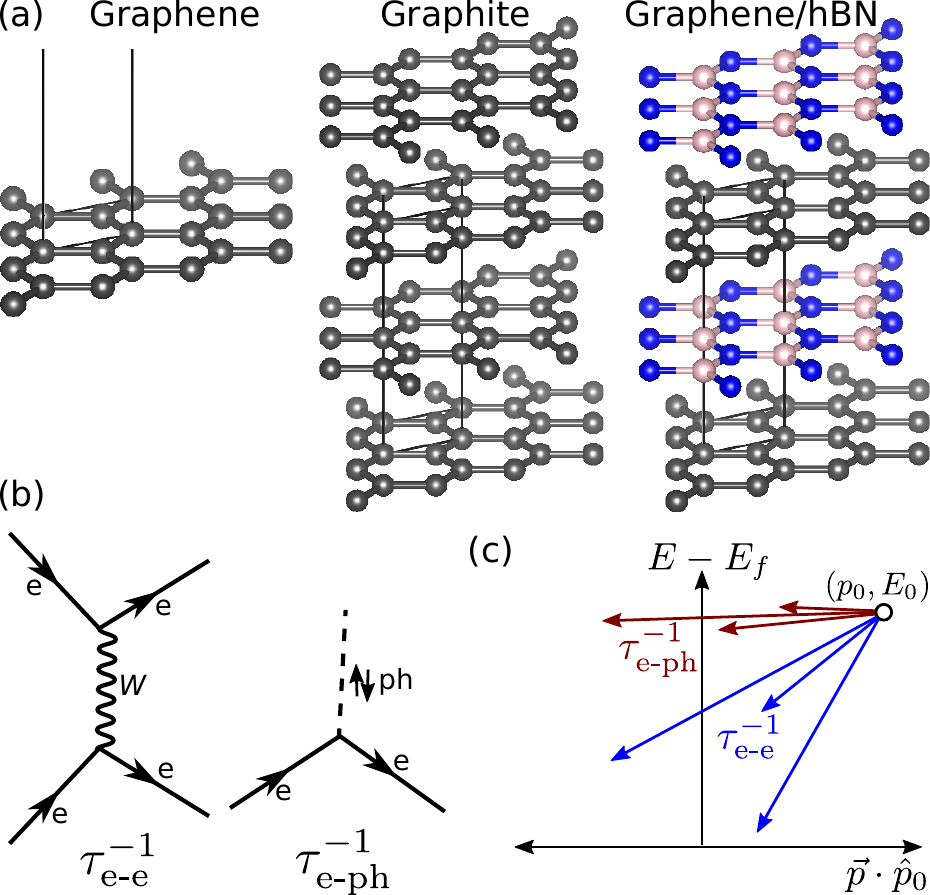}
\caption{(a) Schematic structures of 2D materials and graphene-derived vdW heterostructures
in which we investigate hot carrier relaxation dynamics.
(b) Feynman diagrams for the intrinsic carrier relaxation mechanisms:
electron-phonon (e-ph) scattering and and electron-electron (e-e) scattering.
(c) Roles of the two mechanisms in energy and momentum relaxation:
both processes randomize the momentum component along the original
propagation direction, $\vec{p}\cdot\hat{p}_0$ in a single scatter,
while energy relaxation due to a single scattering event is substantial only
in e-e scattering because the overall energy-scale of phonons
is much smaller than the typical scale of excited electron energies.
\label{fig:schematic}}
\end{figure}

With an \emph{ab initio} framework for calculating optical response and electron-phonon interactions,
we previously evaluated mechanisms of hot carrier generation and relaxation in plasmonic metals,\cite{NatCom,PhononAssisted}
and identified their signatures in ultrafast pump-probe measurements.\cite{TAparameters,TA-analysis}
In particular, the small mean free paths of higher energy carriers helped elucidate the efficiency limits
in plasmonic energy conversion devices and potential strategies to overcome them.\cite{Prineha:2016fk}
Here, we investigate the dynamics of hot carriers in graphene and in graphene-derived vdW heterostructures
to explore their potential for efficient hot carrier extraction. The focus of previous work
in carrier dynamics (that includes electron-phonon coupling in graphene) has primarily been on
low-energy carriers that dominate many properties of interest such as near-equilibrium charge transport.\cite{PhysRevLett.102.076803} From a plasmonic hot carrier devices perspective, we instead focus on first principles calculations of higher energy carrier dynamics.
In particular, we calculate the energy-dependent life times and mean free paths
of hot carriers in free-standing graphene, and additionally to evaluate the role
of inter-layer interactions, in graphite and graphene/hBN (alternating graphene
and hexagonal boron-nitride layers), as shown in  Fig.~\ref{fig:schematic}(a).
Of many possible heterostructures between graphene and hBN, we pick the simplest
configuration which is computationally most tractable with fewest atoms per unit cell,
and captures all the relevant interactions with maximum graphene-hBN interactions.The presence of dopants to shift the Fermi level will
alter the carrier dynamics and a detailed description of this interaction is not the primary focus here.\cite{PhysRevB.94.115208}

The intrinsic carrier relaxation rate in materials is determined by two prominent processes,
electron-phonon (e-ph) scattering and electron-electron (e-e) scattering,
as shown in Fig.~\ref{fig:schematic}(b).
Fermi's Golden rule for e-ph scattering yields the rate\cite{PhononAssisted}
\begin{multline}
\left(\tau\sub{e-ph}^{-1}\right)_{\vec{q}{n}} =
\frac{2\pi}{\hbar} \int\sub{BZ} \frac{\Omega d\vec{q}'}{(2\pi)^3} \sum_{n'\alpha\pm}
	\delta(E_{\vec{q}'n'} - E_{\vec{q}n} \mp \hbar\omega_{\vec{q}'-\vec{q},\alpha})
\\
\times
	\left( n_{\vec{q}'-\vec{q},\alpha} + \frac{1}{2} \mp \left(\frac{1}{2} - f_{\vec{q}'n'}\right)\right)
	\left| g^{\vec{q}'-\vec{q},\alpha}_{\vec{q}'n',\vec{q}n} \right|^2.
\label{eqn:tauInv_ePh}
\end{multline}
Above, electronic states with energies $E_{\vec{q}'n'}$ and Fermi-Dirac occupation factors $f_{\vec{q}n}$
are labeled by wave-vectors $\vec{q},\vec{q}'$ in the Brillouin zone BZ and band indices $n,n'$.
Phonon states with energies $\hbar\omega_{\vec{k}\alpha}$ and Bose-Einstein occupation factors
$n_{\vec{k}\alpha}$ are labelled by wave-vectors $\vec{k}$ ($=\vec{q}'-\vec{q}$
by momentum conservation) and polarization index $\alpha$.
The electron-phonon matrix elements $g^{\vec{q}'-\vec{q},\alpha}_{\vec{q}'n',\vec{q}n}$
couple two electronic states with a phonon mode as in the 3-vertex shown in Fig.~\ref{fig:schematic}(b).

On the other hand, the rate for electron-electron scattering is given by\cite{PhononAssisted,eeLinewidth}
\begin{multline}
\left(\tau\sub{e-e}^{-1}\right)_{\vec{q}{n}} =
\frac{2\pi}{\hbar} \int\sub{BZ} \frac{d\vec{q}'}{(2\pi)^3} \sum_{n'}
\sum_{\vec{G}\vec{G}'}
	\tilde{\rho}_{\vec{q}'n',\vec{q}n}(\vec{G})
	\tilde{\rho}_{\vec{q}'n',\vec{q}n}^\ast(\vec{G}')\\
\times \frac{1}{\pi}\Im\Bigg[ \underbrace{\frac{4\pi e^2}{|\vec{q}'-\vec{q}+\vec{G}|^2}
	\epsilon^{-1}_{\vec{G}\vec{G}'}(\vec{q}'-\vec{q},E_{\vec{q}n}-E_{\vec{q}'n'})
	}_{W_{\vec{G}\vec{G'}}(\vec{q'}-\vec{q}, E_{\vec{q}n}-E_{\vec{q}'n'})} \Bigg].
\label{eqn:tauInv_ee}
\end{multline}
Above, the relevant matrix element for Fermi's golden rule is obtained by the overlap of
the density matrix $\tilde{\rho}_{\vec{q}'n',\vec{q}n}$ between initial and final electronic
wavefunctions, expanded in the plane-wave basis with reciprocal lattice vectors $\vec{G}$,
with the imaginary part of the dynamically screened Coulomb operator
$W_{\vec{G}\vec{G'}}(\vec{q'}-\vec{q}, \hbar\omega)$.
This operator is written in terms of the electronic dielectric function,
which in turn, is derived from the electronic density matrices and energies.
Note that the density matrices contain the lower incoming and outgoing electronic
states in Fig.~\ref{fig:schematic}(b), while $\Im W$ contains the virtual photon
propagator and the upper electronic states.
See Ref.~\citenum{eeLinewidth} for a detailed exposition and
Ref.~\citenum{PhononAssisted} for our implementation details.

Both (\ref{eqn:tauInv_ePh}) for e-ph scattering and (\ref{eqn:tauInv_ee})
for e-e scattering couple electrons with any incoming and outgoing
wavevectors, $\vec{q}$ and $\vec{q}'$, and are therefore capable,
in general, of completely changing the momentum direction in a single
scattering event, as shown in Fig.~\ref{fig:schematic}(c).
The energy conservation in (\ref{eqn:tauInv_ePh}) only couples
incoming and outgoing electronic states differing by a phonon energy.
Since phonon energies are typically $\sim 0.1$~eV or smaller, while relevant
electronic energies are $\sim 1$~eV, e-ph scattering only relaxes
a small fraction of the energy in each scattering event.
In contrast, (\ref{eqn:tauInv_ee}) couples an incoming electronic state
with any outgoing electronic state with a smaller magnitude of
energy,\cite{PhononAssisted} so that e-e scattering relaxes the
hot carrier energy much more efficiently per scattering event,
as indicated in Fig.~\ref{fig:schematic}(c). Coupling and 
scattering with phonon polaritons, hyperbolic modes in hBN and super-collisions
have been considered elsewhere and are not included here.\cite{PhysRevLett.109.056805, Betz:2013dp, PhysRevLett.109.106602}

We perform \emph{ab initio} density-functional theory (DFT) calculations
of the electronic band structure, phonon dispersion relations and
electron-phonon matrix elements used in (\ref{eqn:tauInv_ePh}-\ref{eqn:tauInv_ee}).
We use maximally-localized Wannier functions\cite{MLWFmetal}
to interpolate all the DFT-calculated quantities from
a coarser Brillouin zone mesh ($24 \times 24 \times 4$) to
the much finer meshes ($600 \times 600 \times 12$),
which is critical for accurately resolving the phonon energy scales
in \emph{ab initio} calculations of electron-phonon properties.\cite{PhononAssisted}
For graphene, we use truncated Coulomb potentials\cite{TruncatedEXX}
to isolate periodic images along the third direction,
and set the Brillouin zone sampling to 1 in that direction.
Additionally, we use the plane-wave basis with a kinetic energy cutoff
of 30~Hartrees, norm-conserving pseudopotentials,\cite{SG15}
and the `PBE' generalized-gradient approximation to the
exchange-correlation functional,\cite{PBE} all as implemented
in our open-source DFT software, JDFTx.\cite{JDFTx}
See Ref.~\citenum{PhononAssisted} for further computational
and implementation details.

\begin{figure}
\includegraphics[width=\columnwidth]{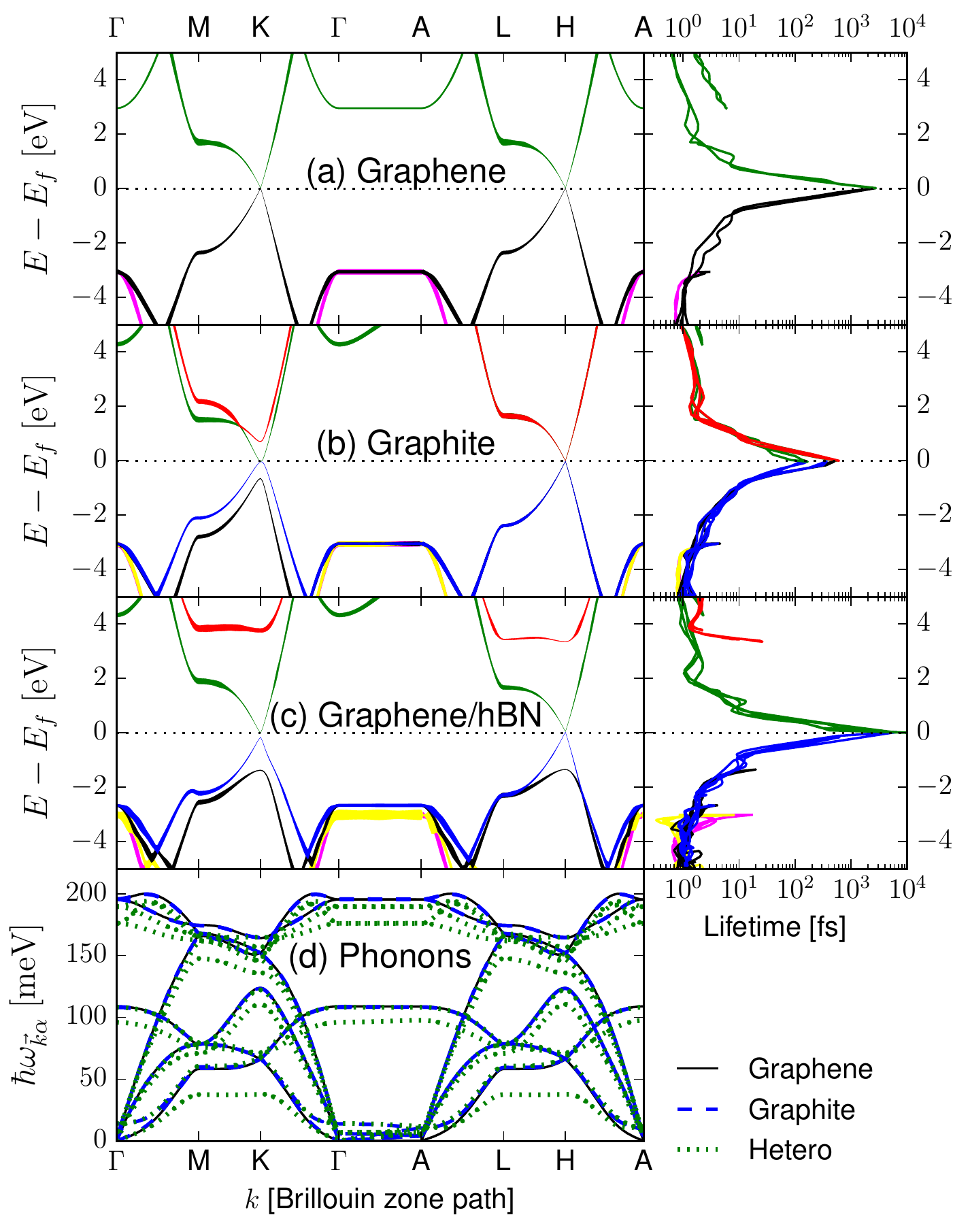}
\caption{Electronic band structure and corresponding carrier lifetimes
of (a) graphene, (b) graphite and (c) the graphene / hBN heterostructure.
The left panels show the band structure with line widths equal
to the calculated carrier linewidth ($\hbar/(2\tau)$),
while the right panels show the total carrier lifetimes ($\tau$),
accounting for e-e and e-ph scattering, on a logarithmic scale.
The $y$-axes (carrier energy) of the left and right panels are matched,
and the band coloring is intended to aid identifying corresponding
features between the band structure and the lifetime panels.
The carrier lifetime typically decreases monotonically with increasing
carrier energy away from the Fermi level, except when new bands begin
at a given energy eg. 4~eV above the Fermi level in (a) and (c).
The maximum lifetime at the Fermi level (limited by e-ph scattering)
is reduced by an order of magnitude in going from graphene (a) to graphite (b),
but is restored when the graphene layers are separated by hBN layers in (c).
Panel (d) compares the calculated phonon bandstructures for the three materials.
\label{fig:bandstructLifetime}}
\end{figure}

Fig.~\ref{fig:bandstructLifetime} shows the calculated electronic bandstructure and
corresponding carrier lifetimes $\tau = (\tau\sub{e-e}^{-1} + \tau\sub{e-ph}^{-1})^{-1}$
for graphene, graphite and graphene/hBN (the heterostructure with alternating graphene
and hexagonal boron nitride layers shown in Fig.~\ref{fig:schematic}(c)).
The shape of the $\tau$ vs $E-E_f$ curves in these materials is
qualitatively similar to that in conventional metals.\cite{PhononAssisted}
For carriers close to the Fermi level, $\tau$ is dominated by e-ph scattering.
With increasing energy, the contribution of e-e scattering increases causing a rapid drop in $\tau$.
The key difference between conventional metals and these materials,
however, is the magnitude of the lifetimes. 
In particular, note that the maximum carrier lifetime in graphene
(near the Fermi level) is approximately 3000~fs, almost two orders
of magnitude larger than the typical maximum lifetimes
of 30-40~fs in noble metals.\cite{PhononAssisted}

Fig.~\ref{fig:bandstructLifetime} also correlates the band structure
for each material with the energy dependence of the carrier lifetimes,
and shows the carrier linewidth ($=\hbar/(2\tau)$) on the band structure as well.
For example, note that near the Dirac point in graphene (at the $K$ point
in the Brillouin zone), the lines narrow to essentially zero width
due to the orders of magnitude increase in the carrier lifetime.
At approximately 2~eV below the (intrinsic) Fermi level,
the carrier lifetimes are smaller by a factor of 2 -- 3
in the $K-M$ segment compared to the $K-\Gamma$ segment,
seen qualitatively in the thicker lines on the bandstructure near $M$,
and read quantitatively off the right panels containing the lifetimes.
Specifically, this decrease in lifetime is due to the flattening of the
energy vs $k$ relation near the $M$ point, which results in a higher
density of states and a greater electron-phonon scattering rate.
In general, the lifetime decreases mostly monotonically with increasing carrier
energy away from the Fermi level, until additional bands become accessible at
higher energies, for example, at approximately 3~eV above the Fermi level. These bands
show up as an additional segment starting at a higher lifetime in the right panel.

In going from graphene to graphite in Fig.~\ref{fig:bandstructLifetime}(b),
the bands crossing at the Dirac point ($K$) pick up a small curvature, and
additional low-lying bands appear approximately 1~eV above and below the Fermi level.
However, at the $H$ point, which is directly above the $K$ point at the
Brillouin zone boundary along the $k_z$ direction, the bands continue to
cross linearly as in the Dirac point in graphene.
(Intuitively, the zone-boundary $k_z$ implies that the wavefunctions are
out of phase between the two graphene layers in the graphite unit cell,
which minimizes the effect of wavefunction overlaps between the two layers
on the band energies at that point.)
Correspondingly, in the lifetime panel, there are two branches:
a higher lifetime branch corresponding to the $H$ point vicinity,
and lower lifetime branch corresponding to the $K$ point.
Even the higher of the two lifetimes at the Fermi level is approximately 600~fs, an
order of magnitude lower than in graphene, because of the increased phase space for
electron-phonon scattering due to the aforementioned band curvature at the $K$ point.

Next, Fig.~\ref{fig:bandstructLifetime}(c) shows that separating graphene layers
with hexagonal boron nitride layers preserves the graphene band structure near the
Dirac point, and restores the high maximum carrier lifetimes of graphene.
This is because, unlike in graphite, the bands of the insulating boron nitride are far
from the Fermi level and do not hybridize with the graphene bands near the Fermi level.
In fact, the maximum carrier lifetimes in the graphene/hBN heterostructure
exceed that of free-standing graphene because adjacent layers clamp down
the out-of-plane vibrations of the atoms, increasing the corresponding
$z$-polarized acoustic (`ZA') phonon frequencies, thereby reducing the corresponding
phonon occupation factors and the electron-phonon scattering rates in (\ref{eqn:tauInv_ePh}).

\begin{figure}
\includegraphics[width=\columnwidth]{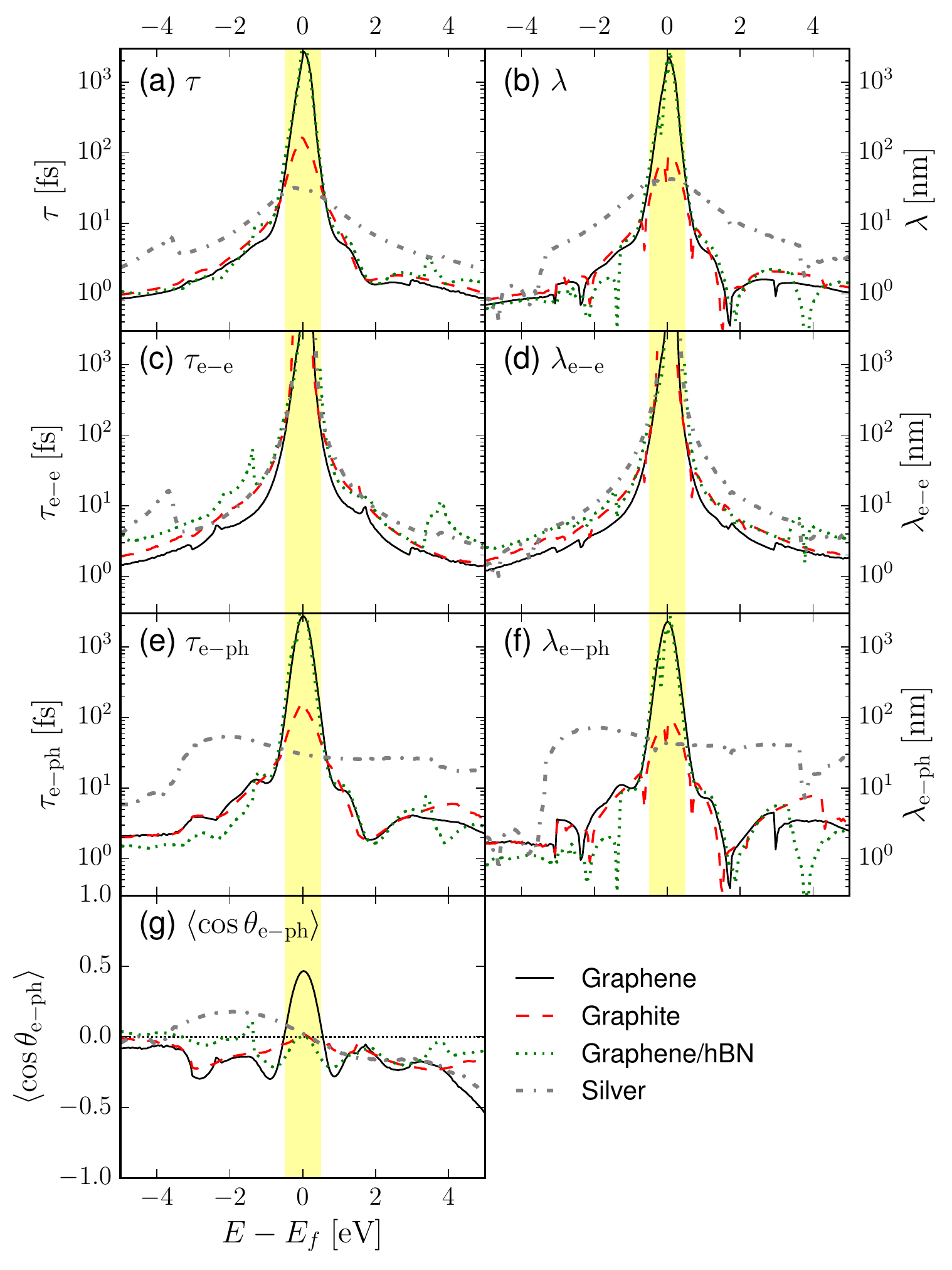}
\caption{Comparison of the average (a) lifetime $\tau$ and (b) mean free path $\lambda$
as a function of carrier energy in the 2D materials and the best-case 3D metal, silver.
Panels (c) and (d) show the corresponding e-e scattering contributions,
while (e) and (f) show the e-ph scattering contributions to $\tau$ and $\lambda$.
Note that the graphene-derived vdW heterostructures
 outperform silver in the highlighted low-energy band
extending $0.5$~eV away from the Fermi level, with a peak Fermi-level lifetime
in graphene and graphene/hBN approximately two orders of magnitude larger than silver.
However, at higher energies, silver exhibits greater lifetimes and transport distances
due to lower electron-phonon scattering rates.
Panel (g) shows the average scattering angle in electron-phonon scattering;
most materials show $\langle\cos\theta\rangle\approx 0$, except graphene
near the Fermi level, where the positive $\langle\cos\theta\rangle$
indicates that carriers retain a portion of their momentum after scattering.
\label{fig:tauLambda}}
\end{figure}

Finally, Fig.~\ref{fig:tauLambda} compares the total values and individual contributions
of e-e and e-ph scattering to the lifetime $\tau$ and mean free path $\lambda = v\tau$
(where the carrier group velocity $v = v_{\vec{q}n} = \nabla_{\vec{q}} E_{\vec{q}n} / \hbar$
is also calculated from the band structure in the Wannier representation).
To make the comparisons between different materials clear, we now plot a single
value for each quantity as a function of energy, averaging out the anisotropic
$k$-dependence previously shown in Fig.~\ref{fig:bandstructLifetime}.
The top two panels, Fig.~\ref{fig:tauLambda}(a,b), show as before that
graphene and graphene/hBN have the highest Fermi level $\tau$ and $\lambda$,
with the corresponding values in graphite one order of magnitude smaller.
In comparison, the peak values for silver are $\tau\sim 30$~fs and $\lambda\sim 40$~nm,
which are an additional order of magnitude smaller than the graphite case.
Therefore, the 2D materials are clearly superior for transport of low-energy carriers,
as is established in previous studies.\cite{PhysRevB.84.075449,Park:2008qd, PhysRevLett.98.076602, PhysRevLett.98.186806}

Importantly, however, the lifetimes and mean free paths of the 2D materials decreases
with carrier energy magnitude much faster than in the case of noble metals like silver.
Consequently, outside the energy window extending 0.5~eV above and below the Fermi level
highlighted in Fig.~\ref{fig:tauLambda}, $\tau$ and $\lambda$ are larger in silver
than they are for any of the 2D materials, by almost a factor of $4-5$ for $3-4$~eV carriers.
Panels (c) and (d) of Fig.~\ref{fig:tauLambda} show that e-e scattering at
higher energies is somewhat comparable between the graphene-derived vdW heterostructures and silver.
Instead, panels (e) and (f) show that the dramatic reduction of
higher-energy $\tau$ in the graphene-derived vdW heterostructures
 is due to increased e-ph scattering,
which in turn is attributable to softer phonon modes and
stronger electron-phonon coupling with the lighter atoms.

Comparisons of the e-e and e-ph contributions between the graphene-derived vdW heterostructures
in Fig.~\ref{fig:tauLambda}(c) and (e) reveal an interesting competition.
The e-e scattering rate at higher carrier energies is highest in graphene,
and is reduced in graphite and graphene/hBN (i.e. $\tau\sub{e-e}$ is increased)
because the neighboring layers contribute in screening the Coulomb interaction
between the electrons and therefore reduce the magnitude of $W$ in (\ref{eqn:tauInv_ee}).
On the other hand, the e-ph scattering rate at higher carrier energies is lowest in graphene,
while the neighboring layers provide additional phonon modes that the electrons can
scatter against thereby increasing the scattering rate (reducing $\tau\sub{e-ph}$).
Therefore, the ideal heterostructure for maximizing the carrier lifetimes
would use spacer layers with a high-dielectric constant for optimal screening,
which are rigid and consist of heavier atoms to reduce the phonon losses.

The shapes and magnitudes of corresponding $\tau$ and $\lambda$ panels are
peripherally very similar, because, coincidentally the typical Fermi velocity in all
the materials considered here is $\sim 1$~nm/fs, but there are important differences.
In particular, note the sharp increase in $\tau\sub{e-e}$ of silver 3.5~eV below
the Fermi level in Fig.~\ref{fig:tauLambda}(c), which is due to additional bands
(the $d$ bands) becoming accessible at that energy.
However, those new bands have much lower group velocities, and therefore this
increase is absent in the corresponding $\lambda\sub{e-e}$ curve for silver
in Fig.~\ref{fig:tauLambda}(d).
Importantly, this implies that it is easier to find longer-lived (high $\tau$)
high-energy carriers, than to find easy-to-collect (high $\lambda$) carriers.

In addition to the scattering rate, an additional factor that affects charge transport
is the change in angle upon scattering, shown in Fig.~\ref{fig:tauLambda}(g).
Here $\langle\cos\theta\rangle = 1$ would imply exclusive forward scattering
and $\langle\cos\theta\rangle = -1$, exclusive back scattering.
Most of the materials exhibit $\langle\cos\theta\rangle$ near zero
over a large energy range, implying that the initial and final electron
momentum directions are uncorrelated on average.
The only exception is graphene near the Dirac point, where
$\langle\cos\theta\rangle \approx 0.5$, indicating that momentum relaxation
is slower by a factor of two than $\tau\sub{e-ph}$ would indicate.
Correspondingly, the mobility of free-standing graphene compared to the other materials
would be a factor of two larger than that inferred by a ratio of $\tau\sub{e-ph}$.

By investigating electron-electron and electron-phonon scattering dynamics
using a parameter-free \emph{ab initio} framework that fully accounts for
detalied electronic structure and phononic properties of the materials,
we have identified an important avenue for 2D heterostructure research:
designing materials with superior high-energy carrier transport properties.
In particular, simultaneously screening electron-electron interactions
and minimizing phonon losses, perhaps by combining light semi-metals (like graphene)
with heavier and rigid dielectrics (such as NbTe\sub{2}),
could enhance hot carrier transport distances.
Simultaneously, the developed computational framework can be applied to
quantitatively analyse ultrafast pump-probe and electron microscopy measurements
and evaluate additional loss mechanisms due to defects and experimental non-idealities.

We acknowledge financial support from NG NEXT at the Northrop Grumman Corporation.
PN acknowledges support from the Harvard University Center for the Environment (HUCE).
Calculations were performed on the BlueGene/Q supercomputer in the
Center for Computational Innovations (CCI) at Rensselaer Polytechnic Institute
and calculations in this work used the National Energy 
Research Scientific Computing Center, a DOE Office of Science User Facility supported by the Office of Science of
the U.S. Department of Energy under Contract No. DE-AC02-05CH11231.

\bibliographystyle{apsrev4-1}
\makeatletter{} 

\end{document}